\documentstyle[11pt,epsfig]{article}

\def\fig#1{Fig.~{\ref{#1}}}
\def\eqn#1{Eq.~(\ref{#1})}
\def\tree{{\rm tree}}

\newskip\humongous \humongous=0pt plus 1000pt minus 1000pt
\def\caja{\mathsurround=0pt}
\def\eqalign#1{\,\vcenter{\openup1\jot \caja
        \ialign{\strut \hfil$\displaystyle{##}$&$
        \displaystyle{{}##}$\hfil\crcr#1\crcr}}\,}
\newif\ifdtup

\newcounter{eqnumber}
\renewcommand{\theeqnumber}{\arabic{eqnumber}}
\def\equn{
\refstepcounter{eqnumber}
\eqno({\rm \theeqnumber})
}

\def\eqn#1{eq.~(\ref{#1})}

\def\fig#1{fig.~{\ref{#1}}}

\def\tr{{\rm tr}}

\newbox\charbox
\newbox\slabox
\def\s#1{{      
        \setbox\charbox=\hbox{$#1$}
        \setbox\slabox=\hbox{$/$}
        \dimen\charbox=\ht\slabox
        \advance\dimen\charbox by -\dp\slabox
        \advance\dimen\charbox by -\ht\charbox
        \advance\dimen\charbox by \dp\charbox
        \divide\dimen\charbox by 2
        \raise-\dimen\charbox\hbox to \wd\charbox{\hss/\hss}
        \llap{$#1$}
}}

\def\spa#1.#2{\left\langle#1\,#2\right\rangle}
\def\spb#1.#2{\left[#1\,#2\right]}
\def\lor#1.#2{\left(#1\,#2\right)}
\def\sand#1.#2.#3{%
  \left\langle\smash{#1}{\vphantom1}\right|{#2}%
  \left|\smash{#3}{\vphantom1}\right\rangle}
\def\sandp#1.#2.#3{%
  \left\langle\smash{#1}{\vphantom1}^{-}\right|{#2}%
  \left|\smash{#3}{\vphantom1}^{+}\right\rangle}
\def\sandpp#1.#2.#3{%
  \left\langle\smash{#1}{\vphantom1}^{+}\right|{#2}%
  \left|\smash{#3}{\vphantom1}^{+}\right\rangle}
\def\sandmm#1.#2.#3{%
  \left\langle\smash{#1}{\vphantom1}^{-}\right|{#2}%
  \left|\smash{#3}{\vphantom1}^{-}\right\rangle}
\def\sandpm#1.#2.#3{%
  \left\langle\smash{#1}{\vphantom1}^{+}\right|{#2}%
  \left|\smash{#3}{\vphantom1}^{-}\right\rangle}
\def\sandmp#1.#2.#3{%
  \left\langle\smash{#1}{\vphantom1}^{-}\right|{#2}%
  \left|\smash{#3}{\vphantom1}^{+}\right\rangle}
\catcode`@=11  

\def\si{\sigma}
\def\Tr{\, {\rm Tr}}
\def\P{{\rm P}}
\def\NP{{\rm NP}}
\def\LC{{\rm LC}}
\def\SC{{\rm SC}}
\def\eps{\epsilon}
\def\Ord{{\cal O}}

\def\I{{\cal I}}
\def\fourperm#1#2#3#4{(#1\,#2\,#3\,#4)}%
\def\susy{{\scriptscriptstyle \rm SUSY}}
\def\pol{\eps}

\begin{document}

\begin{titlepage}

\begin{flushright}
hep-ph/9702424 \hfill UCLA/97/TEP/4\\
February, 1997\\
\end{flushright}

\vskip 2.cm

\begin{center}
{\Large\bf Two-Loop Four-Gluon Amplitudes in N=4 Super-Yang-Mills}
\vskip 2.cm

{\large Z. Bern, J.S.\ Rozowsky and B. Yan}

\vskip 0.5cm

{\it  Department of Physics\\
University of California at Los Angeles\\
Los Angeles,  CA 90095-1547}
\vskip 3cm
\end{center}
\begin{abstract}
Using cutting techniques we obtain the two-loop $N=4$ super-Yang-Mills
helicity amplitudes for four-gluon scattering in terms of scalar
integral functions.  The $N=4$ amplitudes are considerably simpler
than corresponding QCD amplitudes and therefore provide a testing
ground for exploring two-loop amplitudes.  The amplitudes are
constructed directly in terms of gauge invariant quantities and
therefore remain relatively compact throughout the calculation.  We
also present a conjecture for the leading color four-gluon
amplitudes to all orders in the perturbative expansion.
\end{abstract}

\vfill
\end{titlepage}

\baselineskip 16pt
\section{Introduction}

Recent years have seen improvements in the calculational techniques
for one-loop amplitudes \cite{Review}.  Besides calculations of new
amplitudes for phenomenological \cite{FiveParton} purposes, a number
of infinite sequences of amplitudes have been obtained
\cite{AllPlus,Mahlon,SusyFour,SusyOne}.  At two- or higher-loops
there has not been analogous progress as yet, although there is a need
for such calculations in the analysis of experiments. For example,
experiments at LEP are currently sensitive to next-to-next-to leading
order corrections to $Z \rightarrow 3$ jets
\cite{ThreeJetUncertainty}; this requires two-loop amplitudes as an
input.

Before this and other two-loop calculations can be undertaken, a number
of technical issues must be addressed.  It would be useful to have an efficient
technique for obtaining compact analytic expressions for two-loop
amplitudes. Some first steps in developing such a technique have
recently been reported using ideas motivated by string theory
\cite{Long,TwoLoopStrings}. In particular, Reuter, Schmidt and
Schubert \cite{twoloopQEDbeta} have performed an elegant evaluation of
two-loop quantities using a world-line approach.  Kim and Nair
\cite{Nair} have also set up a multi-loop formulation of the recursive
approach \cite{Recursive}. 

Another technical hurdle is that many of the required integrals are
uncalculated \cite{TwoLoopIntegrals}.  The construction of a
numerical program to extract jet cross-sections would also require
significant work; in particular, one would need to extend the known
techniques for handling the infrared divergent corners of phase space
\cite{SliceAndDice}.

In this letter we explore the use of cutting rules
\cite{Cutting,Review} to obtain compact analytical expressions for
two-loop amplitudes.  In particular, we compute the $N\! =\! 4$
supersymmetric four-gluon amplitudes in terms of a basic set of scalar
integral functions.  Amplitudes in $N\! = \!4$ super-Yang-Mills are
particularly simple and are therefore a good starting point.

At tree-level \cite{ParkeTaylor} and one-loop, infinite sequences of
gauge theory amplitudes have been explicitly constructed.  Can one
find such sequences for higher-loop amplitudes?  In this letter we
also present a pattern for the leading-color four-gluon amplitudes to all
orders of perturbation theory consistent with two-particle
cuts. We do not, however, have a proof that the pattern is the
complete answer and it is possible that a more complete analysis
including three- or higher-particle cuts might reveal additional terms.

In the past, cutting rules \cite{Cutting} have been widely used in
field theory and provide powerful constraints on the form of
amplitudes.  In order to construct complete amplitudes one must have
control of the ambiguities and subtractions. At one-loop it has been
shown that for amplitudes satisfying a power counting criterion, such
as supersymmetric ones, one can obtain the complete answer from
four-dimensional cuts and knowledge of the set of integral functions
that can appear in the result \cite{SusyOne}.  For amplitudes not
satisfying the power counting criterion one may unambiguously obtain
the amplitudes by computing the cuts to all orders in the dimensional
regularization parameter \cite{TwoLoopUnitarity,Massive,Review}, since
all terms develop cuts.  At two loops much less is known about the
integral functions.  Nevertheless, for the supersymmetric two-loop
amplitudes considered in this letter one can perform a complete
reconstruction.

\section{Two-loop color decomposition}

At tree and one-loop levels it has proven useful to separate the color
factors from the kinematics \cite{Color}. (The reader may consult
various review articles \cite{ManganoReview,Review} for notation and
further details.)  The same type of color decomposition is also useful
at the two-loop level.  The color decomposed expression for a four-point
amplitude with all particles in the adjoint representation is
$$
\eqalign{
{\cal A}_4^{2 \mbox{-} \rm loop}  (1,2,3,4)
 & = g^6 \hskip -.2 cm  \sum_{\si \in {S_4/Z_4}} 
N_c ^2 \Tr[T^{a_{\si(1)}}T^{a_{\si(2)}}T^{a_{\si(3)}}T^{a_{\si(4)}}]
\, \Bigl( A^{\LC}_{4;1,1}(\si(1), \si(2), \si(3), \si(4)) \cr
& \hskip 4 cm 
     + {1\over N_c^2}
 A^{\SC}_{4;1,1}(\si(1), \si(2), \si(3), \si(4)) \Bigr) \cr
& \hskip 1 cm 
+ g^6 \hskip -.2 cm \sum_{\si \in {S_4/Z_2^3}} N_c 
\Tr[T^{a_{\si(1)}}T^{a_{\si(2)}}] \Tr[T^{a_{\si(3)}}T^{a_{\si(4)}}] \,
A_{4;1,3}(\si(1), \si(2); \si(3), \si(4))\,, \cr}
\equn\label{TwoLoopColor}
$$
where $A_{4;1,1}$ and $A_{4;1,3}$ are `partial amplitudes'.  Our
notation for $A_{n;j,k}$ is that $n$ is the number of external legs,
and $j$ and $k$ label the position at which the color traces are
split; for $n\ge 6$ there can be up to three color traces so for a
generally consistent notation two indices are required.  We have
explicitly broken up the single color trace partial amplitude into
leading color $A_{4;1,1}^{\LC}$ and subleading color $A_{4;1,1}^{\SC}$
pieces.  (By subleading color we mean expressions suppressed in powers
of $N_c$.)  For each external leg we have abbreviated the dependence
on the outgoing external momenta, $k_i$ , and polarizations, $\pol_i$,
by the label $i$. The notation `$S_4/Z_4$' denotes the set of all
permutations of four objects $S_4$, omitting the cyclic
transformations.  The notation `$S_4/Z_2^3$' refers again to the set
of permutations of four objects omitting those permutations which
exchange labels within a single trace or exchange the two traces. That
is, $S_4/Z_2^3 = \{\fourperm1234,\fourperm1324,\fourperm1423\}$.

This decomposition has a straightforward generalization to an
arbitrary number of external legs.  A similar decomposition exists
when some of the particles are in the fundamental representation.  The
partial amplitudes are independently gauge invariant and may therefore
be calculated separately.

\section{Properties of $N=4$ amplitudes}

The high degree of supersymmetry present in $N\! =\! 4$ amplitudes
considerably simplifies their analytic structure.  Supersymmetric
amplitudes with external legs that have either all helicities identical or
all but one identical vanish by a supersymmetry identity \cite{SWI}.
The non-vanishing maximally helicity violating (MHV) amplitudes, where
two of the external helicities are of one type and the rest of the
other type, are especially simple.  At four and five points all the
non-vanishing amplitudes are MHV.

For these reasons four-point $N\! =\! 4$ supersymmetric amplitudes are a
natural starting point for evaluating higher-loop amplitudes.  At
one-loop $N=4$ amplitudes may be considered as pieces of QCD
amplitudes; similarly at two and higher loops we may expect a study of
$N=4$ amplitudes to be of some use for obtaining QCD amplitudes.

The tree-level four-gluon partial amplitudes are 
$$
\eqalign{
& A_4^{\rm tree}(1^\pm, 2^+, 3^+, 4^+) = 0\,, \cr
& A_4^{\rm tree}(1^-, 2^-, 3^+, 4^+) = i {\spa1.2^4 \over 
                             \spa1.2 \spa2.3 \spa3.4 \spa4.1}\,, \cr
& A_4^{\rm tree}(1^-, 2^+, 3^-, 4^+) = i {\spa1.3^4 \over 
                             \spa1.2 \spa2.3 \spa3.4 \spa4.1}\,.  \cr}
\equn\label{TreeAmpls}
$$
These are identical to the ones of standard QCD, since intermediate
scalars or fermions do not propagate at tree level.
The $\pm$ superscripts on the leg labels
represent the helicities of the external gluons.  We have chosen to
represent the amplitudes in terms of the spinor helicity formalism
\cite{SpinorHelicity}, where gluon polarizations are replaced by
spinor inner-products. (The reader may consult review articles
for details \cite{ManganoReview}.)
We use the compact notation
$
\langle k_i^{-} \vert k_j^{+} \rangle \equiv \langle ij \rangle \, , \; 
\langle k_i^{+} \vert k_j^{-} \rangle \equiv [ij] \, ,
$
where $\vert k^{\pm} \rangle$ are massless Weyl spinors labeled by the sign
of their helicities and with normalization $\spa{i}.{j} \spb{j}.{i} 
= 2 k_i \cdot k_j$.

At one-loop the leading color $N\! =\! 4$ partial amplitudes are 
rather simple and given by
$$
A_{4;1}^{1 \mbox{-} \rm loop}(1,2,3,4) = i st \, A_4^{\tree}(1,2,3,4)
 \, \I_4^{1 \mbox{-} \rm loop}(s,t) \, ,
\equn\label{OneLoopNFour}
$$
where
$$
\I_4^{1 \mbox{-} \rm loop}(s,t) =\int {d^{4-2\eps}p \over (2\pi)^{4-2\eps}} \; 
{1\over p^2 (p-k_1)^2 (p-k_1-k_2)^2 (p+k_4)^2 } \, ,
\equn\label{OneLoopBox}
$$
is the one-loop scalar integral. The Mandelstam variables are defined
as $s\equiv (k_1+k_2)^2$ and $t\equiv (k_2+k_3)^2$.  The massless $N
\, = \, 4$ super-multiplet consists of one gluon, four Weyl fermions
and six real scalars, whose contributions are summed over to obtain
the result (\ref{OneLoopNFour}).  The integral (\ref{OneLoopBox}) has
a simple representation in terms of logarithms.  (See for example
ref.~\cite{Long}.)  The remaining subleading color partial amplitudes
may easily be computed in terms of sums of permutations of the
expression (\ref{OneLoopNFour}).  The one-loop $N\!  =\! 4$ four-gluon
amplitude, in $D$ dimensions, was first obtained by Green, Schwarz and
Brink from the low energy limit of superstring theory \cite{GSB}.

Amplitudes, of course, depend on the particular form of dimensional
regularization. The amplitudes~(\ref{OneLoopNFour}) are quoted in the
dimensional reduction scheme \cite{Siegel} or equivalently the
four-dimensional helicity scheme \cite{Long,Review}, which preserve
supersymmetry.  The above expression turns out to be valid to all
orders of of the dimensional regularization parameter $\eps$
\cite{GSB,DimShift}.  Furthermore, when $A_4^{\rm tree}$ is expressed in
terms of formal polarization vectors and spinors instead of
helicities, it is valid in any dimension.  This will be useful
later when we construct two-loop amplitudes.

When using cuts to obtain the $N=4$ gluon amplitudes we also need tree and
one-loop amplitudes with either two external scalars or fermions,
since these particles can also cross the cuts. The maximally helicity
violating amplitudes with two external fermions or scalars may be
obtained directly from the gluon amplitudes using the supersymmetry
identity \cite{SWI,ManganoReview}
$$
\eqalign{
& {\cal A}_n^{\susy} (1_P^-, 2^+, \cdots, j^-, \cdots, (n-1)^+,  n_P^+) = 
              \left({\spa{j}.n \over \spa{j}.1} \right)^{2 - 2 |h_P|}
{\cal A}_n^{\susy} (1^-, 2^+, \cdots, j^-, \cdots, (n-1)^+,  n^+)  \,, \cr}
\equn\label{SusyIdentities}
$$
where leg $j$ is the only negative helicity gluon on the
left-hand-side, $P$ is either a scalar or fermion and $|h_P|$ is the
absolute value of its helicity ($0$ or $1/2$). The legs without subscripts
are taken to be gluons.  Since this identity follows from general
supersymmetry relations, it is valid to all orders in the
perturbative expansion.  

The MHV $N=4$ gluon amplitudes also satisfy the identity \cite{DimShift}
$$
{\cal A}_n(1^+, 2^+, \cdots, i^-, \cdots, j^-, \cdots n^+) = 
{\spa{i}.{j}^4 \over \spa{a}.{b}^4} \,
{\cal A}_n(1^+, 2^+, \cdots, a^-, \cdots, b^-, \cdots n^+) \,,
\equn\label{CyclicIdentity}
$$
where $i$ and $j$ are the only negative helicity legs on the
left-hand-side and $a$ and $b$ are the only negative helicities on the
right-hand-side.  Furthermore, it implies that different cuts
are related by simple relabelings, up to an overall prefactor
of $\spa{i}.{j}^4$, where $i$ and $j$ label the negative helicities
\cite{DimShift}.

In four dimensions the $n$-loop $N=4$ super-Yang-Mills amplitudes are
ultraviolet finite \cite{Mandelstam}. An important ingredient in
Mandelstam's demonstration of finiteness is that for each external leg
a power of external momentum may be extracted from each superspace
Feynman diagram.  More generally, in $D$ dimensions with $N$
supersymmetries (where $N$ is counted in four dimensions), power
counting constraints on the form of the $L>1$ loop effective action
lead to the ultraviolet finiteness condition \cite{SuperSpace},
$$
L < 2 \, {N-1 \over D-4} \,.
\equn\label{Finiteness}
$$
In particular, at two-loops, $N=4$ amplitudes are ultraviolet finite
for $D<7$.

\section{Two-loop cut construction}
\label{TwoLoopCutSection}

%
\begin{figure}[ht]
\begin{center}
\vskip -.7 cm 
\epsfig{file=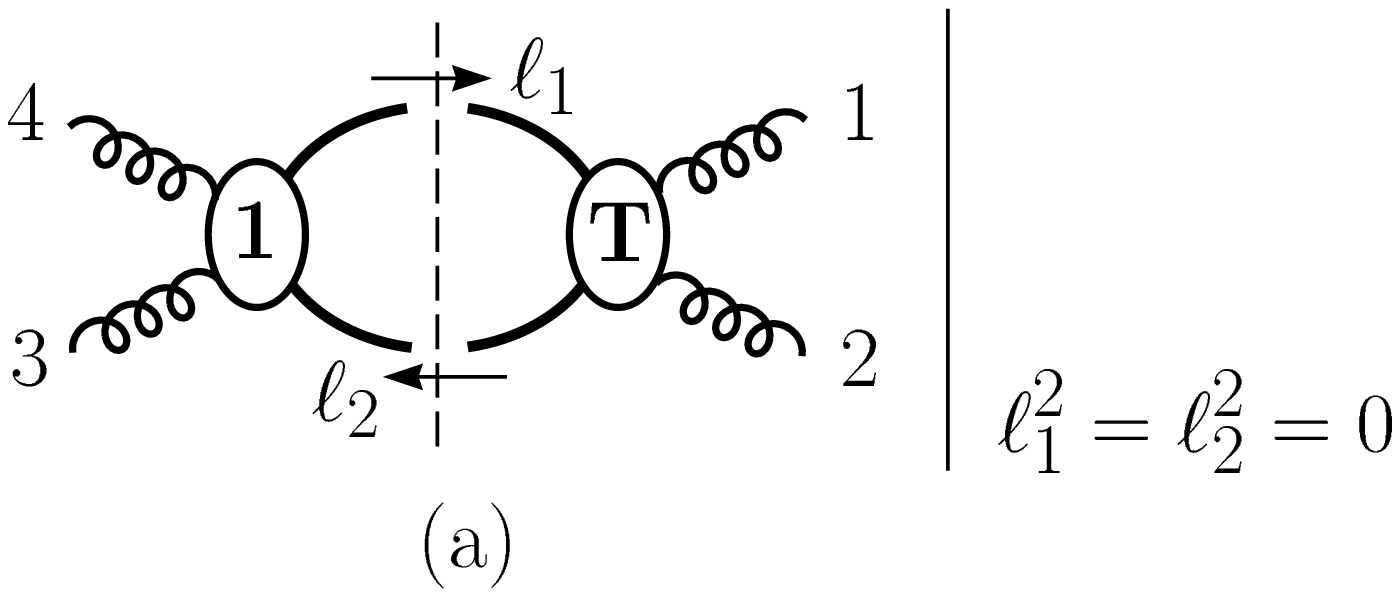,clip=,width=2.3in}
\hskip 1.5 cm
\epsfig{file=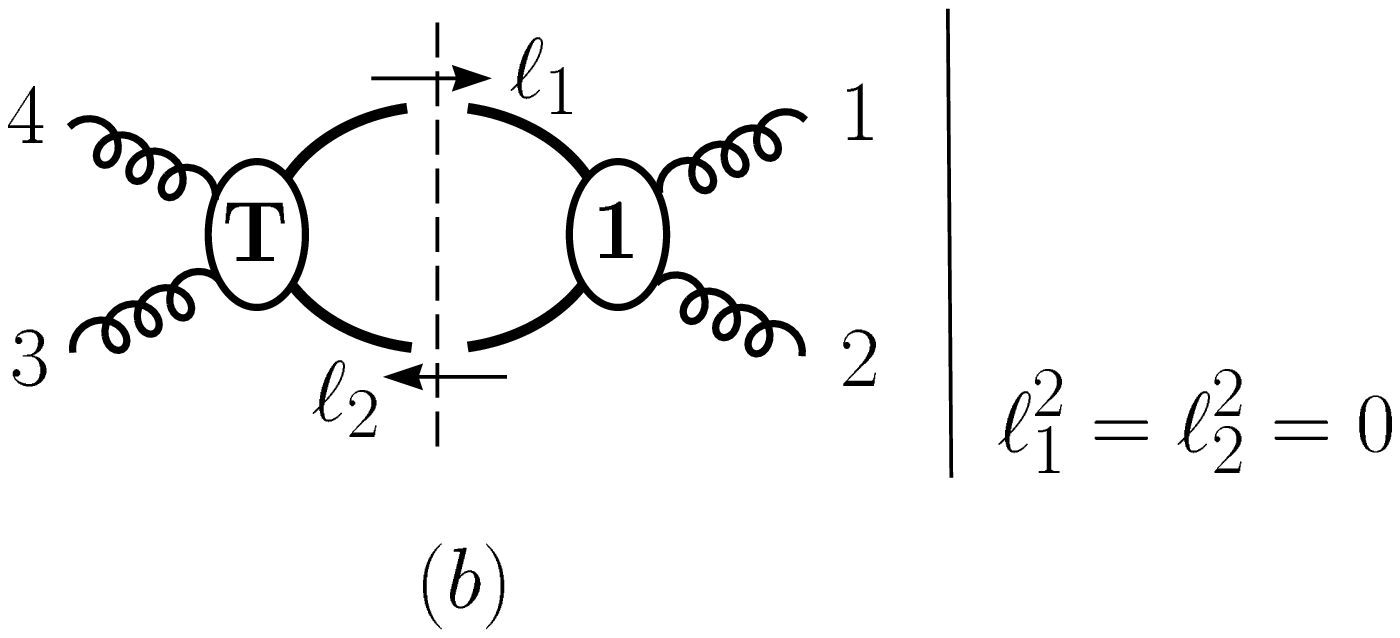,clip=,width=2.3in}
\end{center}
\vskip -.75 cm
\caption[]{
\label{TwoParticleFigure}
\small
The two-particle $s$-channel cut has two contributions: one with the 
four-point one-loop amplitude `1' to the left and the tree amplitude `T' to 
the right (a) and the other being the reverse (b).
}
\end{figure}

First we briefly review the cut construction method
\cite{SusyFour,SusyOne,Review}, extending it to the case of two loops.
Following the discussion in ref.~\cite{Review}, a convenient way to
obtain amplitudes is by considering cuts of unrestricted loop momentum
integrals.  In this way, one may simultaneously construct the
imaginary and associated real parts of the cuts.  Consider for
example, the cuts of the two-loop amplitude ${\cal A}_{4}(1,2,3,4)$.
At two loops one must consider both two- and three-particle cuts.  In
each channel there may be multiple contributing cuts.  For example, in
the $s$ channel there are two two-particle cuts, as depicted in
\fig{TwoParticleFigure}.  The first of these has the explicit
representation
$$
\eqalign{
{\cal A}_{4}^{2 \mbox{-} \rm loop}&(1, 2, 3, 4)
\Bigr|_{\rm cut (a)}  = 
\left. \int\! \sum_{P_1, P_2} {d^{4-2\eps}p\over (2\pi)^{4-2\eps}} \; 
  {i\over \ell_2^2 } \,
 {\cal A}_{4}^{1 \mbox{-} \rm loop} 
    (-\ell_2, 3,4,\ell_1) \,{i\over \ell_1^2} \, 
        {\cal A}_4^{\rm tree} (-\ell_1 ,1,2,\ell_2) 
\right|_{\ell_1^2 = \ell_2^2 =0} \, ,\cr}  
\equn
\label{CutProductDef}
$$
where $\ell_1$ and $\ell_2$ are the momenta of the cut legs and the
sum runs over all particle types (including helicity) $P_1$ and $P_2$
which may propagate across the two cut lines.  We may use the on-shell
conditions $\ell_1^2=0$ and $\ell_2^2=0$ in the integrand (but not on
the cut propagators) since this equation is valid only for those terms
which have explicit $\ell_1$ and $\ell_2$ propagators.  It is
convenient to apply the color decomposition (\ref{TwoLoopColor})
before computing the various cut contributions.

\begin{figure}[ht]
\begin{center}
\vskip -.7 cm
\epsfig{file=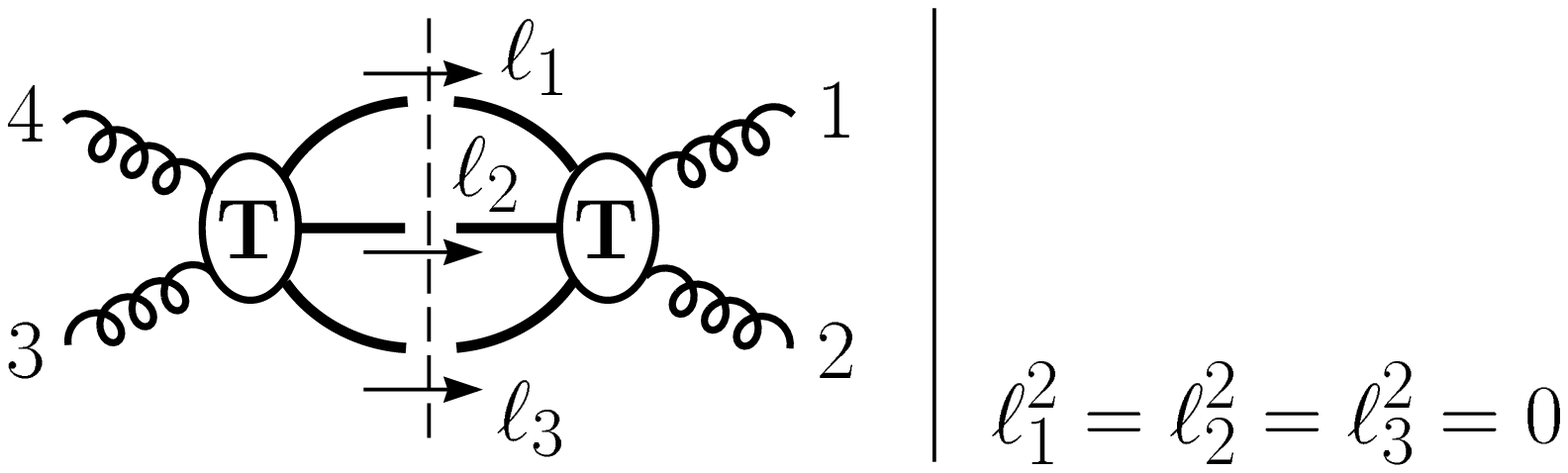,clip=,width=2.8in}
\end{center}
\vskip -.75 cm
\caption[]{
\label{ThreeParticleFigure}
\small
The three-particle $s$-channel cut.}
\end{figure}

The three-particle $s$-channel cut is depicted in
\fig{ThreeParticleFigure}.  The $t$-channel cuts, of course, have a
similar structure to the $s$-channel cuts.  By combining all cuts into
a single function, one obtains the full amplitude.  When combining the
cuts, care must be exercised not to over-count a particular term.

%

As discussed in refs.~\cite{TwoLoopUnitarity,Massive,Review}, by
computing cuts to all orders in the dimensional regularization
parameter, one may perform a complete reconstruction of a massless
loop amplitude.  This follows from dimensional analysis since every
term in an amplitude must have a prefactor of powers of
$(-s_{ij})^{-\eps}$, which necessarily have cuts.  For reasons of
technical simplicity, it is convenient to use helicity amplitudes
which implicitly replace the $(4-2\eps)$-dimensional momenta of the
cut lines by four-dimensional momenta.  However, to do so we must
ensure that errors, especially through $\Ord(\eps^0)$, are not
introduced.  In section~\ref{ValiditySection}, we shall argue that at
least for the $N=4$ four-point amplitudes there are no such errors.

We also find it convenient to perform the cut construction in
components instead of using superfields.  The potential advantage
of a superfield formalism would be that one would simultaneously
include contributions from all particles in a supersymmetry multiplet.
However, for maximally helicity violating amplitudes, the supersymmetry
identities are sufficiently powerful that once the contribution from
one component is known the others immediately follow. A component
formulation is also more natural for extensions to QCD.

\section{Computation of amplitudes}

We now proceed with the explicit construction of the $N=4$ four gluon
partial amplitudes.  First we consider the leading color amplitude
$A_{4;1,1}^{\LC}(1^-, 2^-, 3^+, 4^+)$. In the cut construction we use
the amplitudes (\ref{TreeAmpls}) and (\ref{OneLoopNFour}) as
inputs.  For the $s$-channel cuts depicted in
\fig{TwoParticleFigure}, with the helicity configuration $(1^-, 2^-,
3^+, 4^+)$, only the internal gluon loop contributes in the
configuration where the cut gluons on each side of the cut have
helicity opposite to that of the external legs. For fermion, scalars
or gluons with a different helicity configuration, propagating across
the cut, the amplitudes on both sides of the cut vanish by a
supersymmetry identity. This then gives for \eqn{CutProductDef},
$$
\eqalign{
A_{4;1;1}^{\LC}&(1^-, 2^-, 3^+, 4^+)
	\Bigr|_{\rm cut (a)}\cr
&\hskip -.50cm  = \left. \int\! {d^{4-2\eps}p\over (2\pi)^{4-2\eps}} \; 
 {i\over \ell_2^2 } \,
 A_4^{1  \mbox{-} \rm loop} (-\ell_2^-, 3^+,4^+, \ell_1^-) 
\, {i\over \ell_1^2} \, A_4^{\rm tree} (-\ell_1^+ ,1^-,2^-, \ell_2^+) 
\right|_{\ell_1^2 = \ell_2^2 =0} \cr
&\hskip -.50cm = A_4^\tree \!\left.\left[ \int\!
   {d^{4-2\eps}p\over (2\pi)^{4-2\eps}} \;  s (p + k_4)^2 \, 
  \I_4^{1 \mbox{-} \rm loop}(s, (p+k_4)^2) \, 
{ {\cal N} \over p^2 (p - k_1)^4 (p - k_1 - k_2)^2 (p+k_4)^4} 
\right]\right|_{\ell_1^2 = \ell_2^2 =0} \hskip -.6 cm \,, \cr}
\equn\label{SCutSusyB}
$$
where $\ell_1 = p$ and $\ell_2 = p - k_1 - k_2$
and we have used the amplitudes in eqs.~(\ref{TreeAmpls}) and 
(\ref{OneLoopNFour}). We have also used, for example, 
$1/\spa2.{\ell_2} = - \spb2.{\ell_2}/(p - k_1)^2$
to rationalize the denominators. 
The numerator of the integrand is
$$
\eqalign{
{\cal N} & = \spb{\ell_1}.1 \spa1.4 \spb4.{\ell_1} \spa{\ell_1}.{\ell_2}
             \spb{\ell_2}.3 \spa3.2 \spb2.{\ell_2} \spa{\ell_2}.{\ell_1} \cr
& = \tr_+ [\ell_1 1 4 \ell_1 \ell_2 3 2 \ell_2] \cr
&  = -st \, (p - k_1)^2 (p + k_4)^2 \,, \cr}
\equn
$$
where $\tr_+[\cdots] = {1\over 2} \tr[(1+\gamma_5) \cdots]$.  The
$\gamma_5$ term in the trace does not contribute because a four-point
amplitude has only three independent momenta to contract into the
totally anti-symmetric Levi-Civita tensor. The above sewing algebra
is particularly simple because it is identical to that of the one-loop
case, discussed in ref.~\cite{Review}.

Thus, after canceling numerator factors against propagators and
identifying the remaining integral as a two-loop scalar integral 
we obtain
$$
A_{4;1;1}^{\LC}(1^-, 2^-, 3^+, 4^+)\Bigr|_{\rm cut (a)} = 
-s^2 t \, A_4^\tree(1^-, 2^-, 3^+, 4^+) \, \I_4^{\,\P}(s,t)
\Bigr|_{\rm cut (a)} \,, 
\equn
$$ 
where the two-loop planar scalar double-box integral is 
$$
\I_4^{\,\P}(s,t) = \int
 {d^{4-2\eps}p\over (2\pi)^{4-2\eps}} \;
 {d^{4-2\eps}q\over (2\pi)^{4-2\eps}} \;
 {1\over p^2 \, (p - k_1)^2 \,(p - k_1 - k_2)^2 \,(p + q)^2 q^2 \,
        (q-k_4)^2 \, (q - k_3 - k_4)^2 }\,.
\equn\label{PlanarIntegral}
$$

The evaluation of the two-particle $s$-channel cut in \fig{TwoParticleFigure}b
is similar and gives the same result.  The evaluation of the $t$-channel
two-particle cuts are again similar, but a bit more involved since all
particles in the super-multiplet contribute.  However, after summing over
the contribution of all particles, with the help of the supersymmetry
identities and spinor identities, the integral appearing in the
$t$-channel cut coincides with the one appearing in the
$s$-channel cut in \eqn{SCutSusyB}, but with $s$ and $t$ interchanged.

%
\begin{figure}[ht]
\begin{center}
\vskip -.7 cm
\epsfig{file=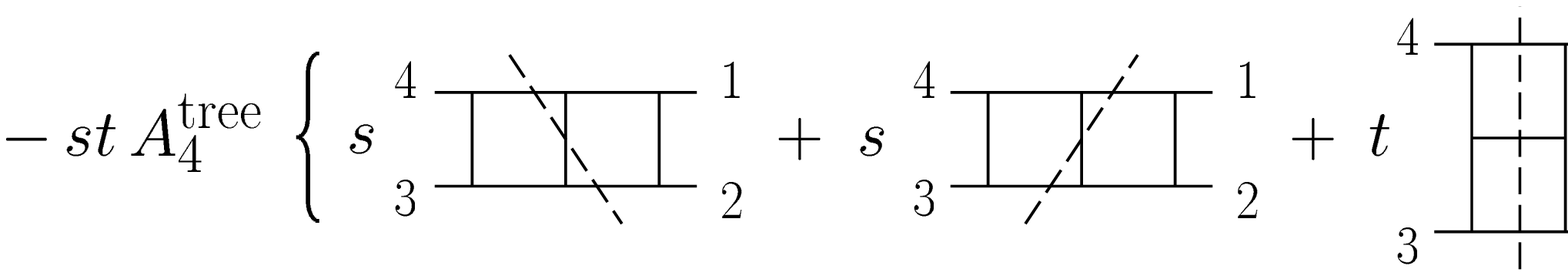,clip=,width=4.2in}
\end{center}
\vskip -.75 cm
\caption[]{
\label{ThreeParticle_ansFigure}
\small The result of evaluating the three-particle $s$-channel cuts in
terms of double-box scalar integrals.  The dashed lines indicate the cuts.}
\end{figure}

The three-particle cuts reproduce the results obtained from the two
particle cuts without adding any new integral functions.  The result
of evaluating the three-particle $s$-channel cut in
\fig{ThreeParticleFigure} is depicted in \fig{ThreeParticle_ansFigure}.
The scalar box integral $\I_4^{\P}(s,t)$ appears twice because it has
two distinct three-particle cuts.  This calculation is more
complicated than the two-particle cut calculation since one must sum
over a total of sixteen intermediate particle and helicity
configurations.  Nevertheless the $N=4$ supersymmetry ensures that 
the various terms combine neatly.

\begin{figure}[ht]
\begin{center}
\vskip -.7 cm
\epsfig{file=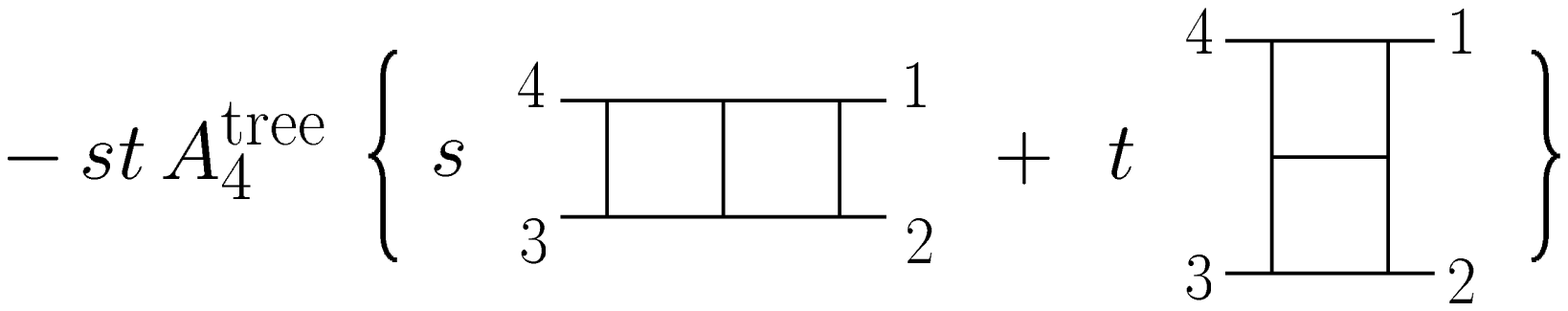,clip=,width=3.3 in}
\end{center}
\vskip -.75 cm
\caption[]{
\label{TwoLoopAmpl_ansFigure}
\small
The result for the leading color two-loop amplitude, corresponding to 
\eqn{LeadingColorResult} }
\end{figure}

Combining all cuts into a single function that has the correct cuts in 
all channels yields the final result
for the leading color $N=4$ four-gluon partial amplitude,
$$
A_{4;1;1}^{\LC}(1^-, 2^-, 3^+, 4^+) = 
-s t \, A_4^\tree(1^-, 2^-, 3^+, 4^+) \, \left( s \, \I_4^{\P}(s,t)
+ t \, \I_4^{\P}(t,s) \right) \,.
\equn\label{LeadingColorResult}
$$
This is depicted in \fig{TwoLoopAmpl_ansFigure}.  We comment that
although substantial progress has been made in the calculation of such
integrals \cite{TwoLoopIntegrals}, the scalar integral appearing in
this amplitude has not yet been evaluated in terms of known
functions.  This amplitude exhibits the cyclic symmetry for the
leading color partial amplitudes (up to an overall $\spa1.2^4$)
implied by the supersymmetry identity (\ref{CyclicIdentity}), as
expected.  We may also use the supersymmetry identity
(\ref{CyclicIdentity}) to immediately obtain the leading color partial
amplitude with non-adjacent negative helicities.

%

The computation of the subleading color pieces is similar to the 
leading color computation, so we only quote the results.
There are two types of subleading color contributions.  Firstly, there
are contributions arising from planar diagrams.  These can be determined
by evaluating the color algebra for the Feynman diagrams and are given
by nothing more than sums of permutations of arguments of the same
functions that appear at leading color.  Secondly, there is a new
contribution characterized by diagrams with non-planar topology.  The
cut construction of the non-planar contributions is quite similar to 
the planar one, with the main difference being the appearance of 
non-planar scalar integrals in the final results.

\begin{figure}[ht]
\begin{center}
\vskip -.7 cm
\epsfig{file=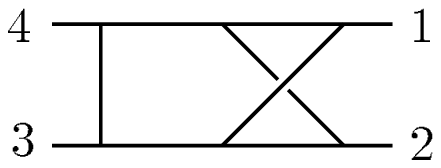,clip=,width=1.5in}
\end{center}
\vskip -.75 cm
\caption[]{
\label{NonPlanarFigure}
\small The non-planar integral function, $\I_4^\NP(s_{12}, s_{23})$,
appearing in the amplitude.  }
\end{figure}

The results for the subleading in $N_c$ partial amplitudes are
$$
\eqalign{
A_{4;1,1}^{\SC}(1,2,3,4) & = 
  2 A_4^{\P}(1,2;3,4) + 2 A_4^{\P}(3,4;2,1)
+ 2 A_4^{\P}(1,4;2,3) + 2 A_4^{\P}(2,3;4,1) \cr
& \hskip 3 cm 
- 4 A_4^{\P}(1,3;2,4) - 4 A_4^{\P}(2,4;3,1) \cr
& \hskip .4 cm 
+ 2 A_4^{\NP}(1;2;3,4) + 2 A_4^{\NP}(3;4;2,1)
+ 2 A_4^{\NP}(1;4;2,3) + 2 A_4^{\NP}(2;3;4,1) \cr
& \hskip 3 cm 
- 4 A_4^{\NP}(1;3;2,4) - 4 A_4^{\NP}(2;4;3,1) \,, \cr
A_{4;1,3}(1;2;3,4) & = 6 A_4^{\P}(1,2;3,4) + 6 A_4^{\P}(1,2;4,3)
+ 4 A_4^{\NP}(1;2;3,4) + 4 A_4^{\NP}(3;4;2,1) \cr
& \hskip .4 cm 
- 2 A_4^{\NP}(1;4;2,3) - 2 A_4^{\NP}(2;3;4,1)
- 2 A_4^{\NP}(1;3;2,4) - 2 A_4^{\NP}(2;4;3,1) \,, \cr}
\equn\label{SubleadingColorResult}
$$
where 
$$
\eqalign{
A_4^{\P}(1,2;3,4) & \equiv  -s_{12}^2 s_{23} \, A_4^{\rm tree}(1,2,3,4)
        \, \I_4^{\P}(s_{12}, s_{23}) \,, \cr
A_4^{\NP}(1;2;3,4) & \equiv  -s_{12}^2 s_{23} \,A_4^{\rm tree}(1,2,3,4)
        \, \I_4^{\NP}(s_{12}, s_{23}) \,, \cr}
\equn
$$
and $s_{ij} = (k_i + k_j)^2$.  The planar scalar integral is defined
in \eqn{PlanarIntegral}, while the non-planar one, depicted in
\fig{NonPlanarFigure}, is
$$
\I_4^{\NP}(s,t) = \int {d^{4-2\eps} p \over (2\pi)^{4-2\eps}} \, 
            {d^{4-2\eps} q \over (2\pi)^{4-2\eps}} \
{1\over p^2\, (p-k_2)^2 \,(p+q)^2 \,(p+q+k_1)^2\,
  q^2 \, (q-k_3)^2 \, (q-k_3-k_4)^2} \,.
\equn
$$

A simple consistency check on the amplitudes (\ref{LeadingColorResult}) and
(\ref{SubleadingColorResult}) is that they satisfy the expected finiteness 
condition (\ref{Finiteness}). 

\section{Validity of cut construction}
\label{ValiditySection}

We now discuss potential errors arising from our use of
four-dimensional helicity amplitudes in the cuts.  Such errors would
not occur if the sewing were done using formal spinors and
polarization vectors.  When using the four-dimensional spinor helicity
formalism potential errors will be of the form,
$$
\int {d^4 p \over (2\pi)^4}  {d^{-2\eps} \mu_p \over (2\pi)^{-2\eps}} \; 
     {d^4 q \over (2\pi)^4}  {d^{-2\eps} \mu_q \over (2\pi)^{-2\eps}} \; 
      {f(p,q, k_i) \times \bigl\{ \mu_p^2, \, \mu_q^2, \, \mu_p\cdot \mu_q,  
            \cdots \bigr\} \over 
      (p^2 - \mu_p^2) (q^2 - \mu_q^2) \cdots } \,,
\equn\label{ErrorTerms}
$$
where we have explicitly separated the loop momenta into four- and
$(-2\eps)$-dimensional parts.  (A discussion of the
$(-2\eps)$-dimensional parts of loop momenta can be found, for
example, in refs.~\cite{Mahlon,Massive}.)  Terms in the integrand
proportional to $(-2\eps)$-dimensional momenta $\mu_p$ or $\mu_q$, may
in principle have been dropped in the above cut construction.  Note
that terms
containing such factors are necessarily suppressed by a power of
$\eps$. However, this suppression may be cancelled if the integrals
contain poles in $\eps$.

The $N\! = \! 4$ amplitudes are, however, special.  We may use the
fact that the algebra associated with the two-particle cuts is
identical at one and two loops. As discussed in
section~\ref{TwoLoopCutSection}, the one-loop amplitudes
(\ref{OneLoopNFour}) are valid in any dimension when expressed in
terms of formal polarizations and spinors.  Since the one-loop
amplitudes are proportional to $A_4^{\rm tree}$, the two-loop sewing
procedure for the two-particle cuts using formal spinors and
polarizations reproduces the one-loop result, except for the
prefactors and the two-loop scalar integral function.  This implies
that all Feynman diagrams with two-particle cuts have been accounted
for exactly.

\begin{figure}[ht]
\begin{center}
\vskip -.7 cm
\epsfig{file=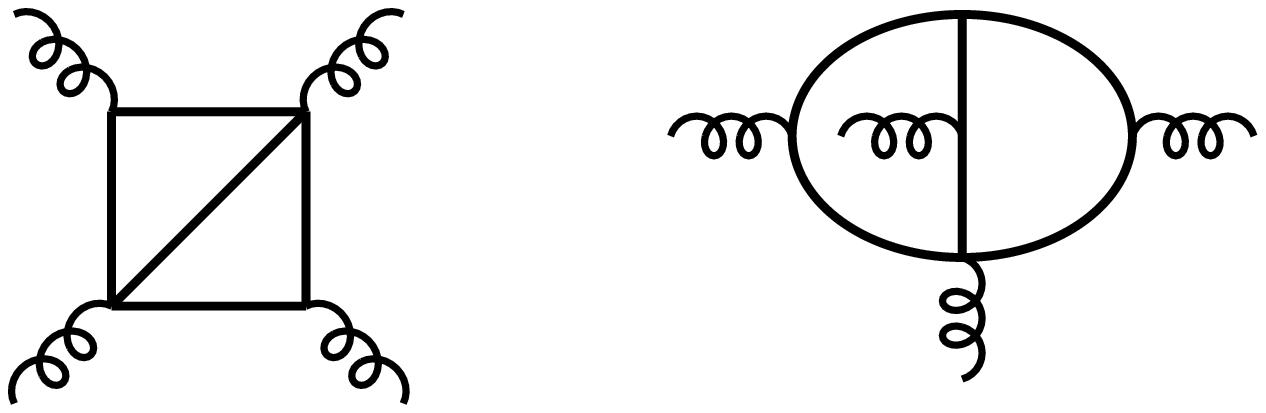,clip=,width=2.4in}
\end{center}
\vskip -.75 cm
\caption[]{
\label{NoTwoParticleCutFigure}
\small Examples of diagrams with no two-particle cuts. The external lines
are gluons, but the internal lines are summed over all states in the 
supermultiplet. }
\end{figure}

This, of course, does not rule out errors of the type in
\eqn{ErrorTerms} from diagrams with no two-particle cuts.  Two
examples of such diagrams are given in \fig{NoTwoParticleCutFigure}.
Here we may appeal to the superspace power counting rule that requires
the extraction of a power of external momentum for each external leg
of a superspace Feynman diagram \cite{Mandelstam}.  After extracting
four powers of external momenta, four-point diagrams with no
two-particle cuts cannot contain any powers of $\mu$ in the numerators
of their integrands since there are no remaining powers of loop momenta
in the non-vanishing diagrams.  This means that there are no
potential errors of the type in \eqn{ErrorTerms} arising from
three-particle cuts.  A more complete discussion of the $\mu$ terms,
especially for theories with less supersymmetries, will be presented
in the future.

\section{Structure of higher loop amplitudes}

Following the same cut construction procedure used for the two-loop
amplitudes, we have found a pattern for the $n$-loop $N\! = \! 4$ 
four-gluon leading color partial amplitudes. 

\begin{figure}[ht]
\begin{center}
\vskip -.7 cm
\epsfig{file=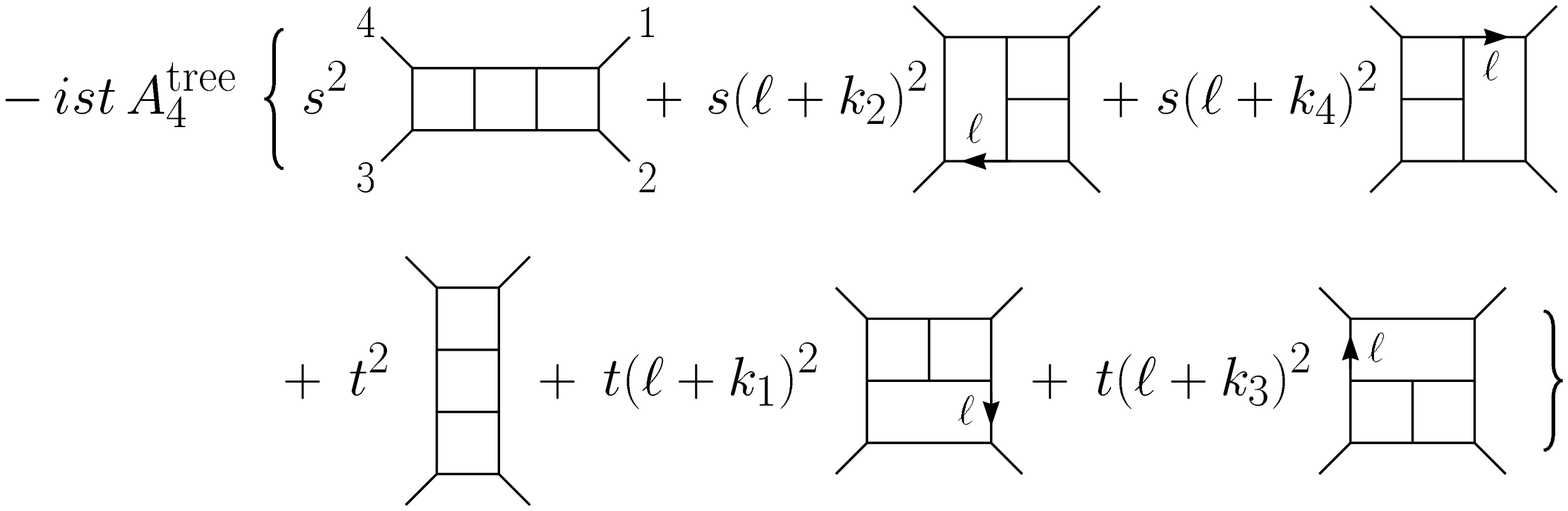,clip=,width=4.5in}
\end{center}
\vskip -.75 cm
\caption[]{
\label{ThreeLoopAmpl_ansFigure}
\small A pictorial representation of the three-loop four-point $N=4$
leading color amplitude. Note the prefactors that involve $\ell$
(where $\ell$ is the internal loop momentum indicated by the arrow in
each term) are part of the integrand.}
\end{figure}

The three-loop leading color partial amplitude is given in
\fig{ThreeLoopAmpl_ansFigure}.  Note that there are one-loop pentagon
sub-diagrams.  This complicates the analysis of the three-particle
cuts since one-loop pentagons can be reduced to sums over box
integrals \cite{Integrals}.  In some cuts it is the box integrals that
appear and in some it is the pentagon; this must be disentangled in
order to identify the form appearing in \fig{ThreeLoopAmpl_ansFigure}.

%

\begin{figure}[ht]
\begin{center}
\vskip -.7 cm
\epsfig{file=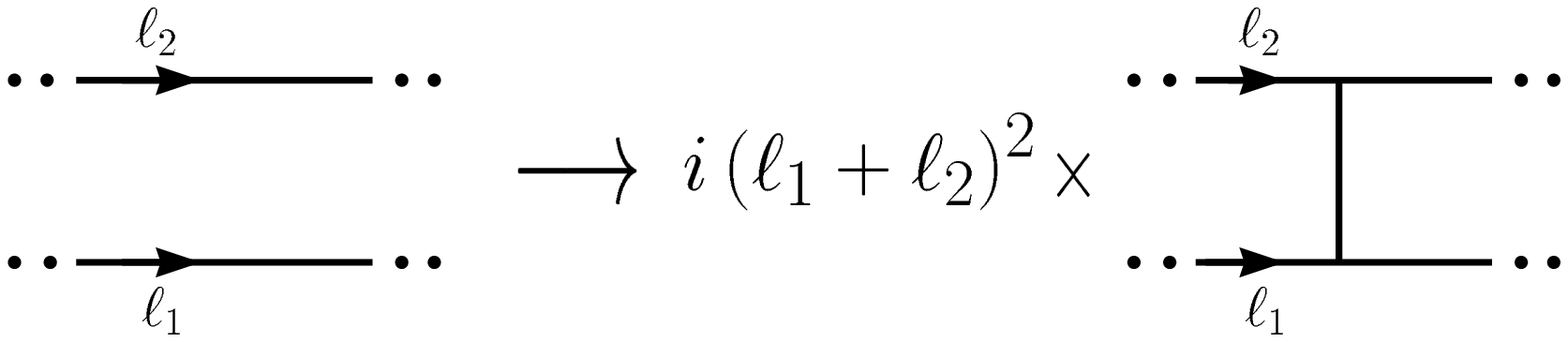,clip=,width=3.0in}
\end{center}
\vskip -.75 cm
\caption[]{
\label{AddLineFigure}
\small
Starting from an $n$-loop integral function we may add an extra line.
The two-lines on the left represent two lines buried in some $n$-loop 
integral.}
\end{figure}

Observing the results for the leading color one-, two- and three-loop
$N=4$ amplitudes one can recognize a pattern which can be used to
construct the $(n+1)$-loop amplitude from the $n$-loop result. The
pattern is that one takes each $n$-loop graph in the $n$-loop
amplitude and generates all the possible $(n+1)$-loop graphs by
inserting a new leg between each possible pair of internal
legs. Diagrams where triangle or bubble subgraphs are created should
not be included.  The new loop momentum including an additional factor
of $i (\ell_1+\ell_2)^2$ in the numerator is integrated over, where
$\ell_1$ and $\ell_2$ are the momenta flowing through each of the legs
to which the new line is joined. (This is depicted in
\fig{AddLineFigure}).  Momentum conservation ensures that it does not
matter on which side of the new line the momentum pair $\ell_1$ and
$\ell_2$ are taken.  Note that no four-point vertices are created by
this procedure. Each distinct $(n+1)$-loop graph should be counted
once, even though they can be generated in multiple ways. The
$(n+1)$-loop amplitude is then the sum of all distinct $(n+1)$-loop
graphs.

%

We have verified that this  pattern is consistent with
two-particle cuts to all loop orders and with four-dimensional three
particle cuts, up to five loops.  The two-particle cuts are not
difficult to check because the same algebra appears at any loop order.
It is, however, possible that there are further terms in the 
amplitudes.  In order to show that nothing has been missed, one would
need to demonstrate that higher order in $\eps$ terms in lower-loop amplitudes
do not cause errors through $\Ord(\eps^0)$.  One would also need to
verify the consistency of higher-particle cuts.  Presumably, a similar
pattern can be found for subleading in color contributions.

\section{Conclusions}

In this letter we have made an initial step in applying the one-loop cut
construction technique \cite{SusyFour,SusyOne} reviewed in
ref.~\cite{Review} to two-loop amplitudes.  As an explicit example we
computed the two-loop $N=4$ supersymmetric four-gluon amplitudes in
terms of a basic set of scalar integrals.  (The amplitudes with two
fermions or scalars and two gluons trivially follow from a
supersymmetry identity \cite{SWI}.)  We also presented a pattern for
the leading-color $N\! = \! 4$ four-gluon amplitudes to all loop
orders consistent with two-particle cuts.

For the two-loop $N\! = \! 4$ amplitudes presented in this letter, we
showed that the cut construction yields the complete amplitude without
any rational function or subtraction ambiguities.  This argument was
based on the fact that the one-loop four-point amplitudes are known in any
dimension and that the amplitudes satisfy a power counting criterion
related to the finiteness of the theory.

A next step, along the lines of this letter, would be to
obtain amplitudes in $N=1,2$ supersymmetric theories as a precursor
to computing QCD amplitudes.  However, before two-loop amplitudes can
be used in phenomenological studies, one would need a numerically
stable method for evaluating the basic integral functions that occur
\cite{TwoLoopIntegrals}. 

\vskip .3 cm 
\noindent
{\large \bf Acknowledgments}

We thank L. Dixon and D.A.~Kosower for comments which greatly improved
this letter.  We also thank D. Dunbar, A. Grant and A.G. Morgan for a
number of helpful comments.  This work was supported by the DOE under
contract DE-FG03-91ER40662 and by the Alfred P. Sloan Foundation under
grant BR-3222.

\end{document}